\documentclass[aip,pof,11pt]{revtex4-2}
\usepackage{subcaption}
\usepackage{graphicx}
\usepackage{dcolumn}
\usepackage{bm}
\usepackage{xcolor, soul}
\usepackage{amsfonts}
\usepackage{booktabs}
\usepackage{siunitx}
\usepackage{ragged2e}
\usepackage[colorlinks,citecolor=red,urlcolor=blue,bookmarks=false,hypertexnames=true]{hyperref} 
\usepackage{float}
\usepackage{float}
\usepackage{xcolor}
\usepackage{array}
\usepackage{setspace}
\usepackage{lipsum}
\usepackage[font=small]{caption}
\newcommand{\BgradB}{$(\mathbf{B} \cdot \nabla) \mathbf{B}$}
\newcommand{\Peclet}{P{\'e}clet }

\newcommand{\HM}[1]{{\color{black}{#1}}}

\newcolumntype{P}[1]{>{\centering\arraybackslash}p{#1}}
\newcolumntype{M}[1]{>{\centering\arraybackslash}m{#1}}
\usepackage{setspace}
\usepackage{amsmath}
\usepackage{parskip}

\begin{document}
\title{Magnetically Assisted Separation of Weakly Magnetic Metal Ions in Porous Media. Part 1: Experiments}

\author{Alwell Nwachukwu}
\affiliation{Department of Chemical and Biomedical Engineering, FAMU-FSU College of Engineering, Tallahassee, FL, 32310, USA}
\affiliation{Center for Rare Earths, Critical Minerals, and Industrial Byproducts, National High Magnetic Field Laboratory, Tallahassee, FL 32310, USA}

\author{Muhammad Garba}
\affiliation{Department of Chemical and Biomedical Engineering, FAMU-FSU College of Engineering, Tallahassee, FL, 32310, USA}
\affiliation{Center for Rare Earths, Critical Minerals, and Industrial Byproducts, National High Magnetic Field Laboratory, Tallahassee, FL 32310, USA}
\author{Jamel Ali}
\affiliation{Department of Chemical and Biomedical Engineering, FAMU-FSU College of Engineering, Tallahassee, FL, 32310, USA}
\affiliation{Center for Rare Earths, Critical Minerals, and Industrial Byproducts, National High Magnetic Field Laboratory, Tallahassee, FL 32310, USA}
\author{Theo Siegrist}
\affiliation{Department of Chemical and Biomedical Engineering, FAMU-FSU College of Engineering, Tallahassee, FL, 32310, USA}
\affiliation{Center for Rare Earths, Critical Minerals, and Industrial Byproducts, National High Magnetic Field Laboratory, Tallahassee, FL 32310, USA}
\author{Munir Humayun}
\affiliation{Center for Rare Earths, Critical Minerals, and Industrial Byproducts, National High Magnetic Field Laboratory, Tallahassee, FL 32310, USA}
\affiliation{Department of Earth, Ocean and Atmospheric Science, Florida State University, Tallahassee, FL 32304, USA.}

\author{Hadi Mohammadigoushki}
\thanks{Corresponding author}\email{hadi.moham@eng.famu.fsu.edu}
\affiliation{Department of Chemical and Biomedical Engineering, FAMU-FSU College of Engineering, Tallahassee, FL, 32310, USA}
\affiliation{Center for Rare Earths, Critical Minerals, and Industrial Byproducts, National High Magnetic Field Laboratory, Tallahassee, FL 32310, USA}

\date{\today}

\begin{abstract}

We report experiments on the magnetophoresis of paramagnetic (MnCl$_{2}$) and diamagnetic (ZnCl$_{2}$) metal ions in porous media under the influence of a non-uniform magnetic field generated by a permanent magnet. Experiments were carried out in a range of initial ion concentrations (1-100 mM), porous media particle sizes (63 $\mu$m and 500 $\mu$m), and varying mixture ratios of metal ion concentrations. For single-ion magnetophoresis, paramagnetic MnCl\(_{2}\) migrated toward the magnet surface, with an enrichment of approximately 2–4\% near regions of high magnetic field. Conversely, diamagnetic ZnCl\(_{2}\) moved away from regions of highest magnetic field gradients, with depletion levels of 0.5–1.8\% relative to the initial concentration. Our results demonstrate that magnetophoresis is directly proportional to porous media particle size, increasing with larger particle sizes, a trend attributed to the reduced drag forces experienced by the ions in media with larger particles. Interestingly, in binary mixtures, both MnCl\(_{2}\) and ZnCl\(_{2}\) migrated toward regions of highest magnetic field, contrary to their individual behaviors. The magnetophoretic effect of MnCl\(_{2}\) was diminished with increasing concentrations of ZnCl\(_{2}\), indicating interactions between the two ions. These findings suggest that both metal ions undergo field-induced cluster formation, with cluster sizes in the micrometer range, in both single and binary ion systems. In binary mixtures, the two ions appear to interact, potentially forming mixed clusters containing both MnCl\(_{2}\) and ZnCl\(_{2}\).

\end{abstract}
\maketitle
\section{Introduction}

As global consumption of electronic devices accelerates, electronic waste (e-waste) has become one of the fastest-growing waste streams worldwide. According to the World Health Organization (WHO), 62 million tonnes were generated in 2022, with projections reaching 74 million tonnes by 2030~\cite{holuszko2021electronic,who2022}. This rapid growth poses significant environmental and health challenges due to the hazardous materials contained within e-waste. From smartphones and laptops to electric vehicles and renewable energy systems, modern electronics rely heavily on critical metals that are embedded in complex matrices and often co-exist with toxic substances, making their recovery both essential and technically challenging~\cite{ibrahim2023challenges}. Conventional methods for extracting and separating these metals, typically involve solvent-solvent extraction, high-temperature smelting or acid leaching, are energy-intensive, generate toxic byproducts, and pose serious environmental and health risks~\cite{song2022sustainable}. As demand for these metals intensifies due to the clean energy transition and technological innovation, there is an urgent need to develop sustainable, low-impact separation and recovery technologies. Many critical metals found in e-waste exhibit distinct magnetic properties, including paramagnetism and diamagnetism, offering a largely untapped opportunity to explore magnetically assisted separation strategies~\cite{iranmanesh2017magnetic,rizos2024understanding,benhal25}. \par 

{Compared to conventional methods like solvent-solvent extraction, this technique offer advantages in simplicity, lower energy consumption, and reduced environmental impact~\cite{iranmanesh2017magnetic}. Magnetically assisted separation is applied across various important fields, including medicine and diagnostics \cite{saias2009,chen2013chip}, molecular biology~\cite{hirschbein1982, safarik2009, berensmeier2006, inglis2006}, wastewater treatment~\cite{ambashta2010, borghi2011, mariani2010, kemsheadl1985}, food production, and pharmaceutical processes\cite{kemsheadl1985,oliveira2010high}. } In the presence of a non-uniform external magnetic field, neutrally charged magnetic particles in a solution experience a net magnetic force, which can be written as 
\cite{schadewald2010,butcher2022,greene1971,leventis2005,svendsen2020,svoboda2004}:\begin{equation}
\label{eq:Kelvin}
    \mathbf{F}_{m} = \frac{4 \pi}{3}\frac{\Delta \chi_m b^3}{\mu_{0}} \nabla(c \mathbf{B} \cdot \mathbf{B})
   \end{equation}
where $\Delta \chi_{m}$, $b$, $c$, $\mu_{0}$, and $\mathbf{B}$ are the difference between the molar magnetic susceptibility of the magnetic particle and the surrounding medium, the radius of the magnetic material, the concentration of the magnetic material in the surrounding medium, the permeability of the vacuum, and the magnetic flux density, respectively. \par 

Recently, we have demonstrated, both experimentally and through modeling, that magnetically assisted separation techniques offer an efficient and environmentally friendly alternative for separating mixture components based on their magnetic properties~\cite{benhal25,rassolov2024magnetophoresis,Rassolov25,khan25}. In these experiments, magnetic metal ions are dispersed in a fluidic environment, enabling selective manipulation and recovery based on their magnetic susceptibility. In a fluidic environment, a magnetic particle is subject to various external forces beyond the magnetic force, including thermal diffusion from Brownian motion, viscous drag, gravitational, and inertial forces \cite{benhal25,yavuz2009,oberteuffer1974,shliomis1974,rodrigues2017,kolczyk2019,svoboda2004,lei2017,leong2016}. For magnetophoresis (i.e., the transport of magnetic particles under a magnetic field) to occur, the magnetic force must overcome these competing forces. However, the interplay among these forces is inherently complex, making it challenging to fully characterize magnetophoretic behavior of a material. As a foundational step toward unraveling this complexity, it is crucial to first isolate and examine the interaction between magnetic forces and thermal diffusion. To experimentally investigate this interplay, magnetophoresis studies can be performed in porous media, offering a controlled environment to systematically evaluate the competing effects of magnetic and thermal diffusion forces.\par 

Within porous media, thermal diffusion serves as the primary opposing force to the magnetic force. Consequently, the key dimensionless parameter for evaluating the extent and feasibility of magnetophoresis is the magnetic \Peclet number, defined as: 
\begin{equation}
    {Pe_{m}} = \frac{\mathbf{F}_{m}}{\mathbf{F}_{D}} = \frac{4\pi \Delta \chi_m b^4 \nabla(c \mathbf{B}\cdot \mathbf{B})}{3\mu_0 k_{B} T}
    \label{pem}
\end{equation}
where $k_B$ is the Boltzmann constant, and $T$ is temperature. \HM{Note that in the above equation, we have approximated the Brownian force as $\mathbf{F}_{D} \approx k_B T/b$\cite{yavuz2006}}. Thus, for magnetophoresis to be effective in porous media, the magnetic \Peclet number must exceed unity. Although there are several studies on magnetophoresis of superparamagnetic microparticles and nanoparticles in bulk fluid\cite{yavuz2006,pamme2004,leong2020unified,leong2015magnetophoresis}, magnetophoresis studies of particles in porous media are scarce. Existing research on magnetophoresis in porous media is limited to metal ions. Several studies have reported on magnetophoresis of various metal ions under the influence of an external magnetic field gradient (see Table~(\ref{tab:Litsummary}) for a summary of experimental results)~\cite{rassolov2024magnetophoresis,fujiwara2001,fujiwara2004,fujiwara2006,franczak2016}. \par 

Fujiwara and co-workers \cite{fujiwara2001,fujiwara2004,fujiwara2006} investigated transport of various transition metal ions on a silica gel support, using a magnetic field gradient of 410 kOe$^{2}$ cm$^{-1}$ supplied by a superconducting magnet, and observed successful magnetophoresis. They observed noticeable migration of paramagnetic metal ions towards the highest magnetic field noted by significantly high magnetophoresis velocity ($v_{mig}$). Given that metal ions have a hydration radius of approximately 6 $\r{A}$\cite{persson2024structure,marcus1988ionic} and considerably lower magnetic susceptibilities compared to superparamagnetic micro or nanoparticles, it is expected that the diffusive force will dominate the magnetic force, as indicated by the calculated magnetic \Peclet numbers shown in Table~(\ref{tab:Litsummary}) that are much smaller than unity. Fujiwara and colleagues hypothesized that, under an external magnetic field, metal ions aggregate into clusters approximately 1 $\mu$m in size, explaining their experimental results. They further observed that metal ions with higher magnetic susceptibility and initial concentrations exhibited greater magnetophoresis velocity, while diamagnetic ions showed no detectable movement~\cite{fujiwara2004}. Franczak et al. demonstrated the magnetophoresis of other metal ions in gelatin under a magnetic field gradient of 14.5 and 189 T$^{2}$ m$^{-1}$ \cite{franczak2016}. They observed an enrichment of 0.8$\%$ for paramagnetic metal ions and a depletion of 0.6$\%$ for diamagnetic ions in the region of high magnetic field strength. As in Fujiwara's experiments, the results of Franczak et al. also suggest that metal ions should have formed clusters of larger complexes under magnetic field gradients.\par 

\begin{table}[hthp]
    \centering
    \begin{tabular}{ccccccc}
    \toprule
        Reference & Porous Media &  $\chi$ & Metal ion & Pe$_{m}$ & $ \Delta$C/C$_{o}$~ & $v_{mig}$\\
        &   & 10$^{6}$ [cm$^{3}$/mol] &  &  & $[\%]$& [mm/hr]\\
    \midrule
        Fujiwara et al.~\cite{fujiwara2001,fujiwara2004,fujiwara2006}  & Silica Gel &14200& MnCl$_{2}$ &  2$\times$10$^{-6}$ & - &5.8\\
         &  &14600  & FeCl$_{3}$ &  2$\times$10$^{-6}$ &  - &7.1\\
         &   &9500 & CoCl$_{2}$ &   1$\times$10$^{-6}$ &  -&5.4\\
         &   &4200 &NiCl$_{2}$ &   6$\times$10$^{-7}$ & -&2.6\\
         &   &6200 &CrCl$_{3}$ &   8$\times$10$^{-7}$ & -&4.5\\
         &   & 1500&CuCl$_{2}$ & 2$\times$10$^{-7}$ & -&1\\
         &  & -10 &ZnCl$_{2}$ &  4$\times$10$^{-10}$ & -&0\\
         &   &-24& AgCl & 1$\times$10$^{-9}$ & -&0\\
         &   &-22 &CdCl$_{2}$ & 1$\times$10$^{-9}$ & -&0\\
         Franczak et al.\cite {franczak2016} & Gelatin & 89600 &DyCl$_{3}$ & 5$\times$10$^{-9}$ &  0.8&-\\
          &  & 27930&GdCl$_{3}$ &  2$\times$10$^{-9}$ & 0.6 &-\\
          &  & -993&YCl$_{3}$ &  6$\times$10$^{-11}$ &  0.6 &-\\
    \bottomrule
    \end{tabular}
     \centering
    \caption{Summary of previous literature on magnetophoresis of metal ions in porous media.}
    \label{tab:Litsummary}
\end{table}

While there are reports of successful magnetophoresis of metal ions in porous media under a magnetic field gradient, several inconsistencies still exist across different studies. Specifically, Fujiwara and colleagues \cite{fujiwara2001,fujiwara2004,fujiwara2006}, reported larger migration distance for metal ions with larger magnetic susceptibility and initial concentration. Under similar conditions, Chie and colleagues~\cite{chie2003} reported that in a mixture containing cobalt chloride and iron chloride (CoCl$_{2}$/FeCl$_{2}$) spotted on silica gel, CoCl$_{2}$ migrated a greater distance than FeCl$_{2}$, despite CoCl$_{2}$ having a lower magnetic susceptibility. The migration of CoCl$_{2}$ farther than FeCl$_{2}$ is unexpected, as the magnetic \Peclet number for FeCl$_{2}$ is higher. The discrepancy between magnetic susceptibility and movement is presumably due to adsorption of the metal ions to the silica gel. Silica gels have a large surface area that facilitates the active adsorption of metal ions \cite{tran1999,iamamoto1989}. Unlike Fujiwara and co-workers' findings, Franczak and colleagues~\cite{franczak2016} found no correlation between magnetic susceptibility and magnetophoresis behavior. They observed slightly different concentration changes for both strongly paramagnetic dysprosium chloride DyCl$_{3}$ and weakly diamagnetic yttrium chloride YCl$_{3}$ ions that otherwise differ significantly in their magnetic susceptibility and, therefore, in the magnetic \Peclet number~\cite{franczak2016}. The adsorption of metal ions to the porous media could also impact the results of the latter study. Using numerical simulations, we have recently demonstrated that the adsorption mechanism may compete with magnetophoresis, potentially influencing the observed experimental results~\cite{rassolov2024magnetophoresis}. Previous studies have not systematically examined or ruled out the impact of metal ion adsorption on the magnetophoresis of metal ions in porous media. \par 

The discrepancies in previously reported data, the lack of a definitive correlation between magnetic susceptibility and magnetophoresis, and the intricate interplay between magnetophoresis and metal ion adsorption in porous media collectively highlight a critical knowledge gap. A deeper, more systematic investigation is essential to unravel the fundamental mechanisms governing the magnetophoresis of metal ions in porous environments. In this first study of a two-part investigation, we experimentally examine the magnetophoresis of transition metal ions in a model porous medium while explicitly eliminating adsorption effects. MnCl\(_2\) is chosen as the paramagnetic ion and ZnCl\(_2\) as the diamagnetic reference due to their distinct magnetic properties. Among transition-metal ions, MnCl\(_2\) exhibits one of the strongest paramagnetic responses, while ZnCl\(_2\) demonstrates pronounced diamagnetic behavior, making them ideal for assessing magnetophoresis in porous media. We will systematically investigate the magnetophoresis of these metal ions in porous media composed of silica gel, exploring a broad range of silica gel particle sizes and initial ion concentrations. Importantly, we also investigate, for the first time, the feasibility of magnetically assisted separation of these metal ions from an initially mixed solution. In the second part of this study, we will develop a comprehensive multi-physics theoretical framework to analyze and interpret the experimental findings presented here.

\section{Materials and Methods}
\label{Methods}
Metal ions, specifically manganese (II) chloride tetrahydrate (MnCl$_{2}$ $\cdot$ 4H${2}$O) and zinc chloride (ZnCl$_{2}$), were procured from Sigma-Aldrich and ThermoFisher Scientific, respectively, and used as received. To prepare ion solutions, these ions are dissolved in deionized water to concentrations of 1 mM, 10 mM, 100 mM. Several studies have reported that lower pH levels decrease the adsorption capacity of silica gel and increase the removal efficiency of metal ions from the silica gel \cite{guo2018,li2014,repo2011}. Therefore, to eliminate the adsorption of metal ions onto the porous media, the acidity of the solution is adjusted to a pH of 3 by adding 10 mM hydrochloric acid (HCl) to all prepared solutions. The acidic pH protonates and decreases the charge on the metal ions and surfaces, minimizing the electrostatic attractions that cause adsorption. For the porous media, we used three different systems. First, we used commercially available polydisperse silica particles (Sigma-Aldrich, 75 - 650 $\mu$m, 800 m$^{2}$/g; our measured particle size distribution is shown in Fig.~S1 of the supplementary materials). In addition, commercially available silica gel, typically offered in a range of particle size distributions, was filtered through size-rated sieves to obtain silica gel with two distinct particle sizes: 63 $\mu$m and 500 $\mu$m (see Fig.~S2 of the supplementary material for microscope images of different sized silica particles.) The silica gel bed is prepared by saturating dry silica beads with the desired metal ion solution. \par 

\begin{figure}[ht]
   
     \centering
    \begin{subfigure}{0.33\textwidth}
        \centering
        \includegraphics[width=\textwidth]{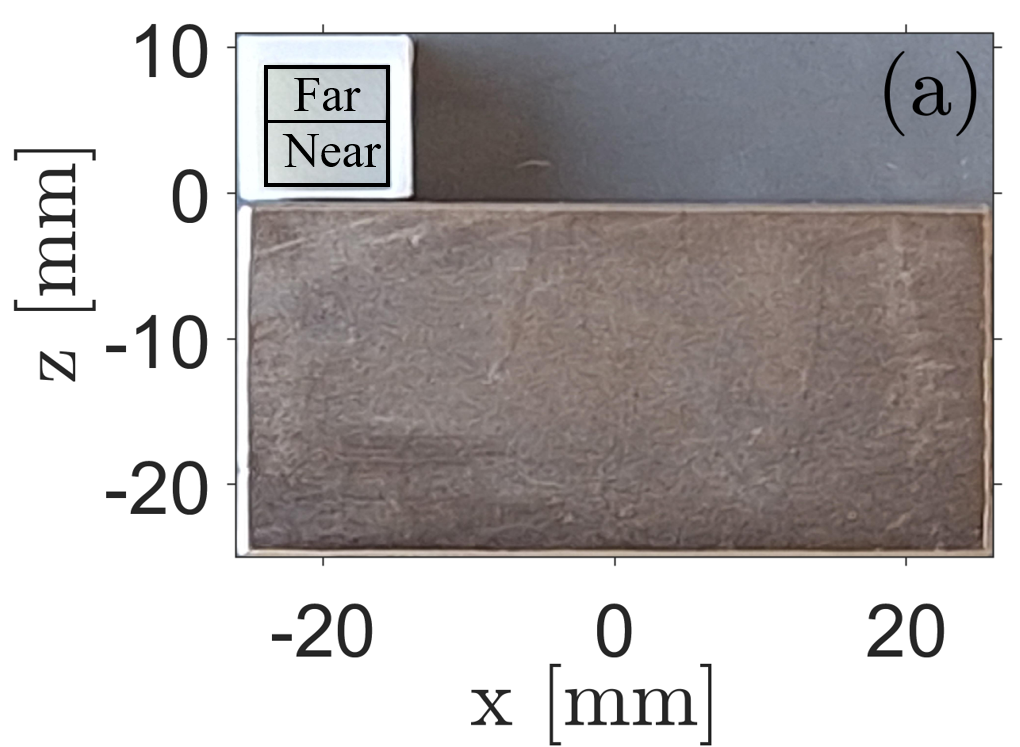}
        \label{magnetplacement}
    \end{subfigure}
    \begin{subfigure}{0.35\textwidth}
        \centering
        \includegraphics[width=\textwidth]{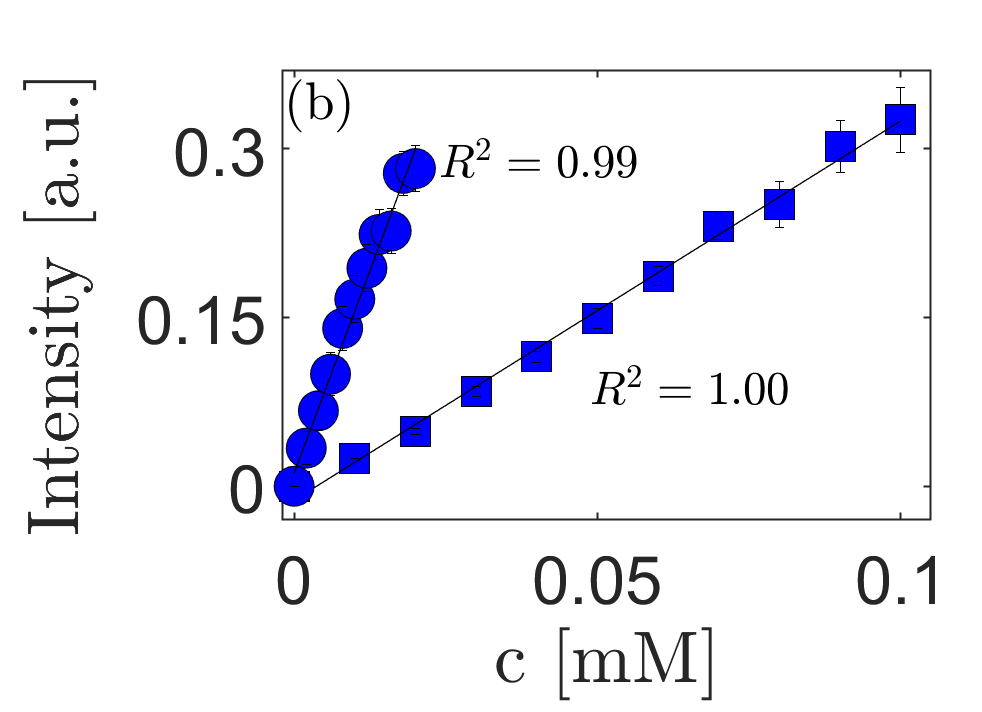}
        \label{UVstdcurve}
    \end{subfigure}
    \begin{subfigure}{0.3\textwidth}
        \centering
        \includegraphics[width=\textwidth]{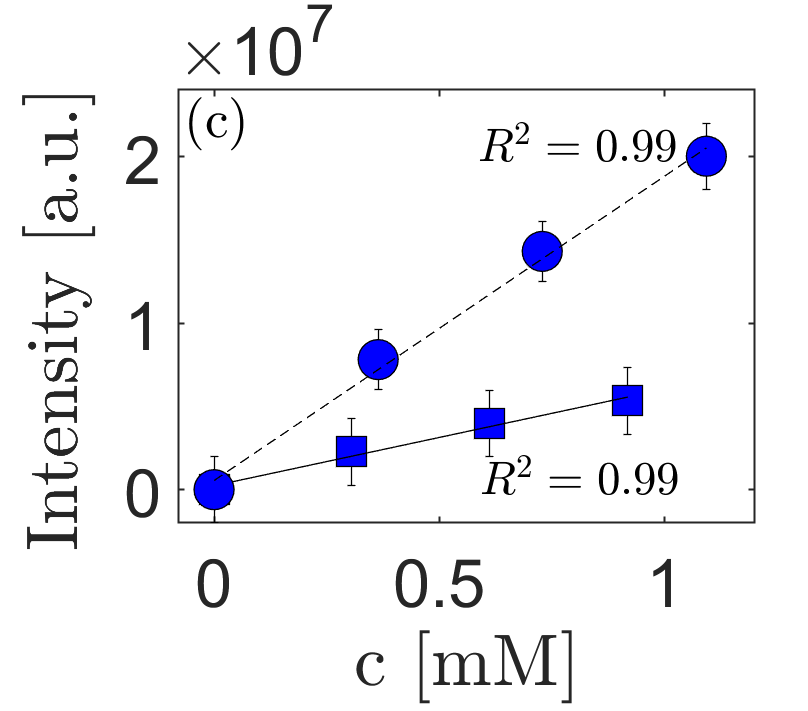}
        \label{ICPstdcurve}
    \end{subfigure}
\caption{(a) a 2D snapshot showing the silica gel containing cell on the permanent magnet. The standard calibration curve for MnCl$_{2}$ (circle) and ZnCl$_{2}$ (square) metal ions obtained using (b) UV spectrophotometer, and (c) the ICP-OES.}
    \label{fig:stdcurves}
\end{figure}

The silica gel bed is positioned within an experimental cell and placed at the edge of a 2$\times$1$\times$1 inch NdFeB permanent magnet to maximize exposure to the magnetic field gradient. Fig.~\ref{fig:stdcurves}(a) illustrates the magnetophoresis experimental setup, featuring the experimental cell containing the silica gel bed positioned directly on the surface of the permanent magnet. The cell is exposed to a magnetic field for up to 72 hours. Magnetic field intensity was experimentally measured using a M3D-2A-PORT portable magnetic field mapping system. \par

Following magnetic field exposure, the gel is carefully sliced into two equal sections. The section closest to the magnet is termed the "near" slice, while the section furthest from the magnet is termed the "far" slice. The gel slices are then centrifuged at 6000 rpm for 60 minutes to separate the silica particles from the solution. The ion concentration in each gel slice is subsequently determined using calibration curves (Fig.~\ref{fig:stdcurves}(b,c)), which were generated using two well-established analytical methods. For solutions containing single metal ions, ion concentrations are determined primarily by colorimetry, with a UV-Vis spectrophotometer (Agilent Cary 3500). This method relies on light absorbance in the presence of 0.08 mM xylenol orange (Sigma-Aldrich), and applies the Beer–Lambert law \cite{swinehart1962beer}. For binary mixture experiments, Inductively Coupled Plasma Optical Emission Spectrometry (ICP-OES; Agilent 5800) is used. For the single-ion solution experiments, a calibration curve is generated by recording the intensity as a function of known ion concentrations. For binary mixture experiments, calibration curves are prepared using a commercially available multi-element ICP standard in a 2–5\% hydrochloric acid matrix (obtained from high-purity standards (HPS)).\par 

\section{Results}
\subsection{Static Magnetic Field}
\begin{figure}[H]
   \centering
      \begin{subfigure}{0.38\textwidth}
        \centering
        \includegraphics[width=\textwidth]{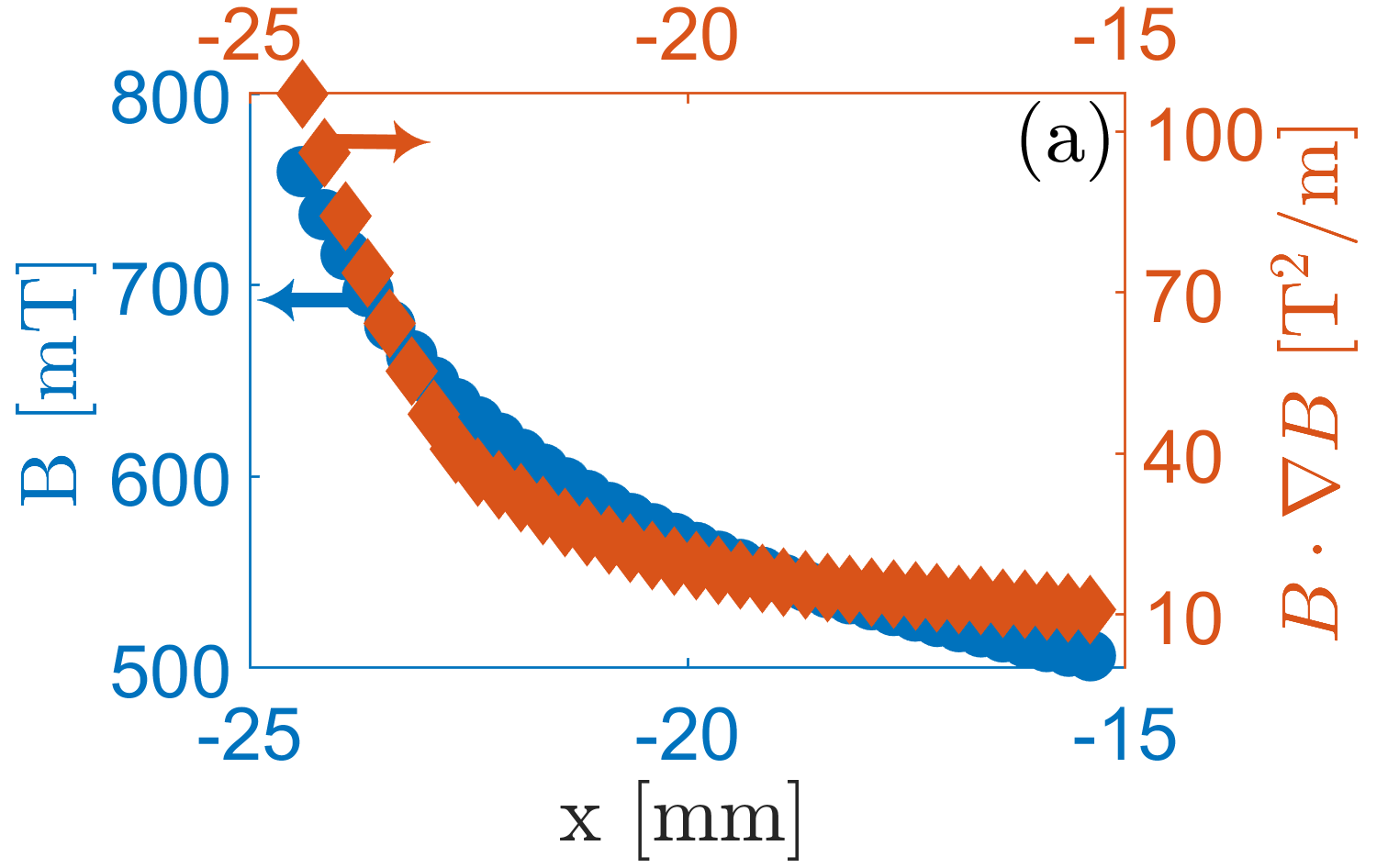}
    \end{subfigure}
    \begin{subfigure}{0.29\textwidth}
        \centering
     \includegraphics[width=\textwidth]{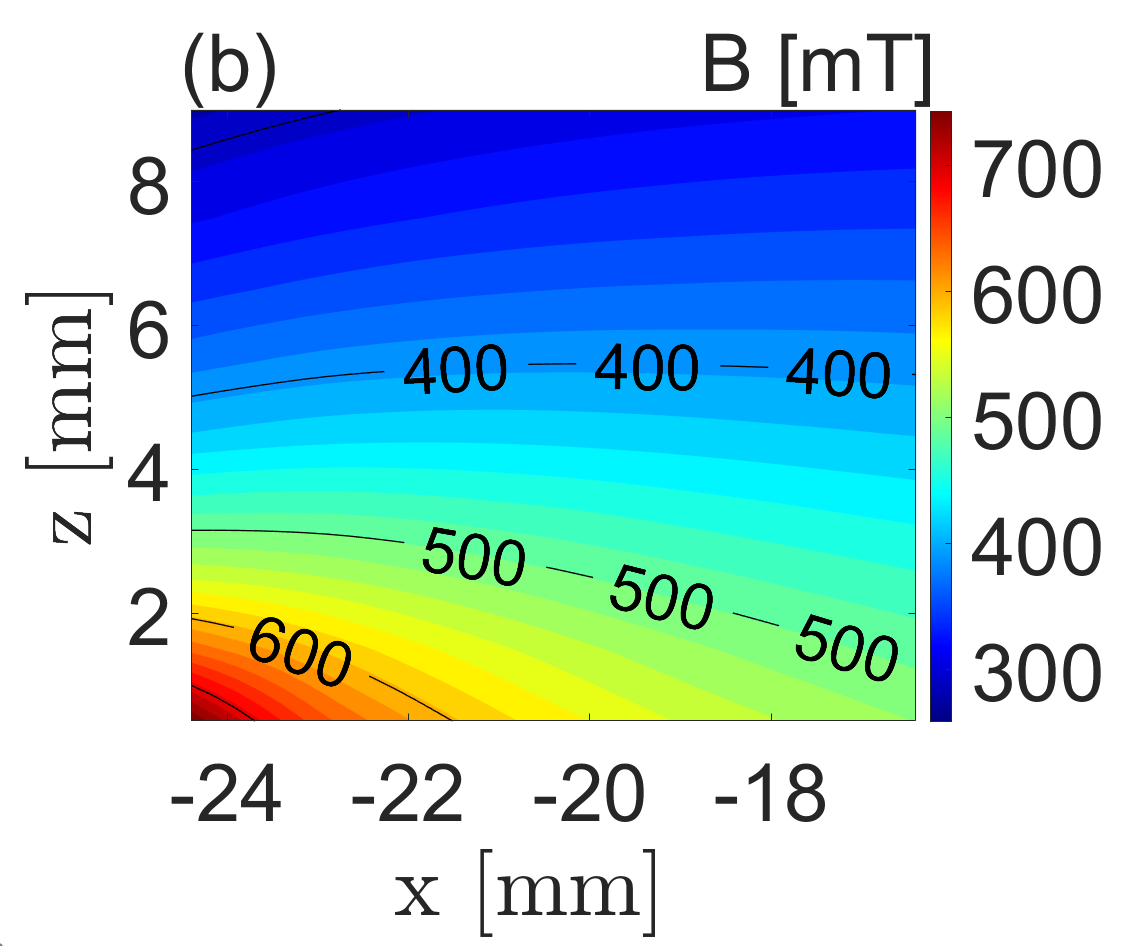}
    \end{subfigure}
    \begin{subfigure}{0.29\textwidth}
        \centering
        \includegraphics[width=\textwidth]{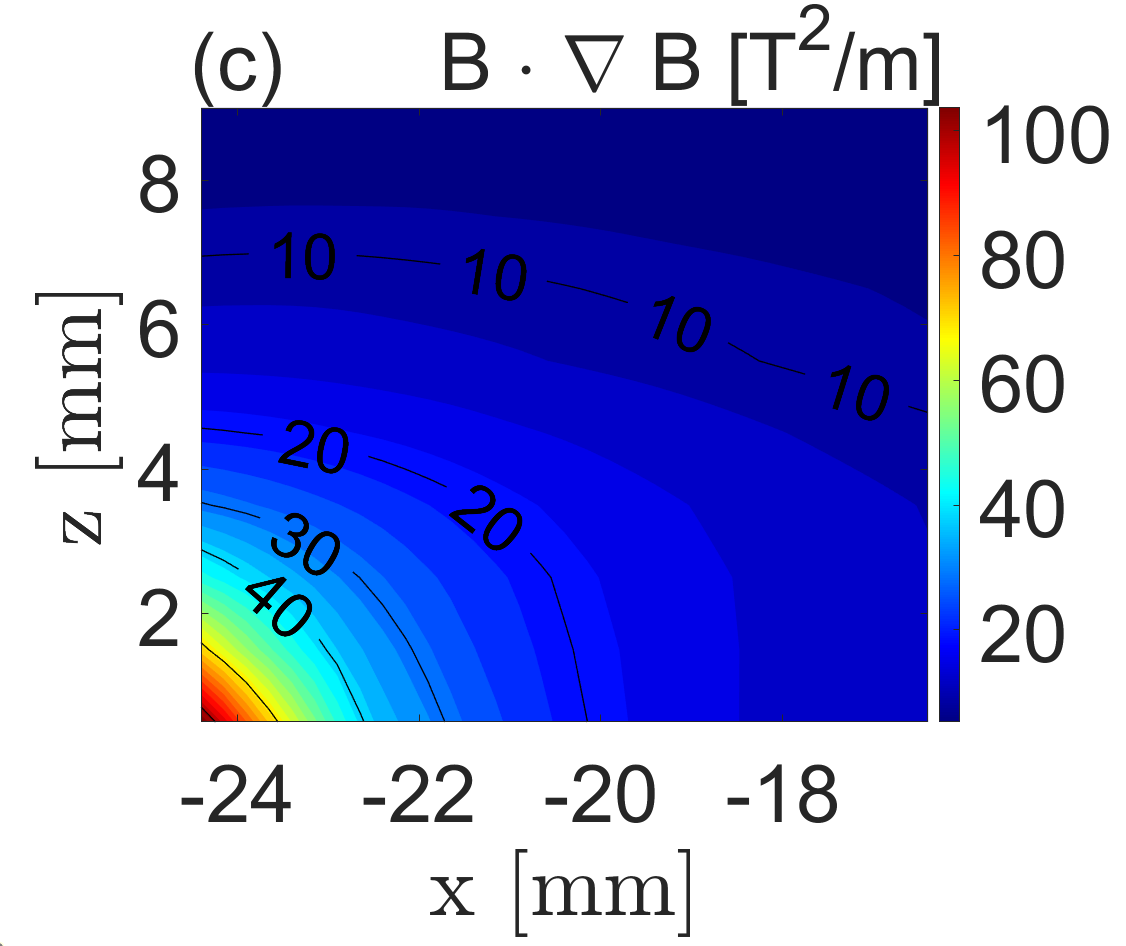}
        
    \end{subfigure}
    \caption{(a) Magnetic flux density (primary y-axis) as a function of location along the magnet surface at z = 0, and magnetic field gradients (secondary y-axis) at z = 0 along the magnet surface. The 2D map of magnetic flux density (b), and magnetic field gradients (c), in the measuring cell used for magnetophoresis study. }
    \label{2DBgradient}
\end{figure}

\HM{The geometry of the magnet, together with the relative positioning of the porous medium, directly determines the magnetic field gradient, $\mathbf{B}\cdot\nabla \mathbf{B}$, experienced by the metal ions. For the present study, our experiments were designed for a scenario that maximizes this gradient and hence maximizes the separation. To this end, we mapped the magnetic field distribution around the largest permanent magnet available to us noted above and identified the corner region as the location of the strongest gradients (see Fig.~S3 in the supplementary materials). Because the field strength decays with increasing distance from the magnet surface, the porous medium was positioned at this corner to ensure that it was subjected to the maximum attainable magnetic force.}
Fig.~\ref{2DBgradient}(a) displays the magnetic flux density $\mathbf{B}$ (primary y-axis) and the magnetic flux density gradient (secondary y-axis) as a function of the distance from one end of the magnet. The magnitude of the magnetic flux density is calculated $|\mathbf{B}| = (|\mathbf{B}_x^{2}| + |\mathbf{B}_z^{2}|)^{1/2}$. Notably, the magnetic field gradient reaches its maximum at the edge of the magnet. Therefore, the experimental cell is positioned at the corner of the permanent magnet for maximum magnetic force. Fig.~\ref{2DBgradient}(b) shows the two-dimensional map of the magnetic flux density magnitude inside the experimental cell. In addition, Fig.~\ref{2DBgradient}(c) shows the calculated magnetic field gradients within the experimental cell domain. The magnetic field gradient is maximum at the bottom left corner of the cell and decreases as we move away from the surface of the magnet. The maximum field gradient in this experimental setup $|\mathbf{B}\cdot \nabla \mathbf{B}|_{max} \approx $ 100 [T$^2$/m]. 

\subsection{Magnetophoresis of single metal ions in polydisperse silica gel} 
\begin{figure}[htp]
    \centering
    \includegraphics[width=\textwidth]{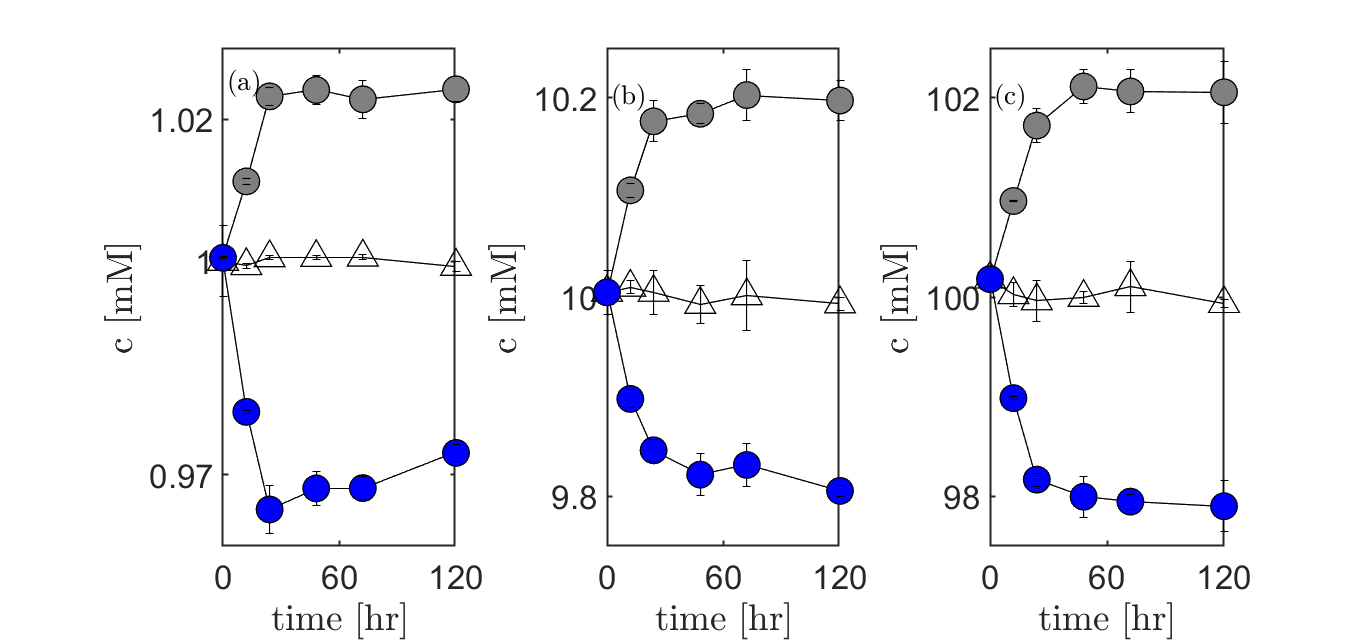}
      \caption{Temporal evolution of paramagnetic MnCl$_{2}$ salt concentration (circle) in polydisperse silica gel induced by a NdFeB permanent magnet ($\mathbf{B}$ = 0.74 T, \BgradB$|_{max}$ = 100 T$^{2}$/m) for initial concentrations of (a) 1 mM (b) 10 mM, and (c) 100 mM. Markers are presented as follows: blue is far section, gray is near section, and triangle denotes the control experiments with $\mathbf{B}$ = 0. }
    \label{fig:MnResults}
\end{figure}

Following the mapping of the magnetic field within the measuring cell, magnetophoresis experiments were performed with solutions containing a single metal ion. Fig.~\ref{fig:MnResults}(a-c) shows the temporal evolution of concentration of the paramagnetic MnCl\(_2\) ion in both the far and near slices for various initial concentrations. Over time, the MnCl\(_2\) ion becomes increasingly enriched near the surface of the magnet, where the magnetic field is strongest. After approximately 24 hours, the ion concentration levels off. A moderate concentration enrichment of \((C - C_0) / C_0 \approx 2 - 2.3\%\) is observed in the near section of the silica gel for all initial concentrations. Here $C_0$ is the initial concentration of metal ion in the porous media. Also included in Fig.~\ref{fig:MnResults}(a-c) are control experiments, where a similar setup was used in the absence of a magnetic field (i.e., \(|\mathbf{B} \cdot \nabla \mathbf{B}| = 0\)). These control experiments show that adsorption or evaporation do not affect the extracted ion concentration, confirming that the observed concentration changes in the presence of the external magnetic field are due to the magnetic field gradients.

\begin{figure}[hthp]
    \centering
    \includegraphics[width=\textwidth]{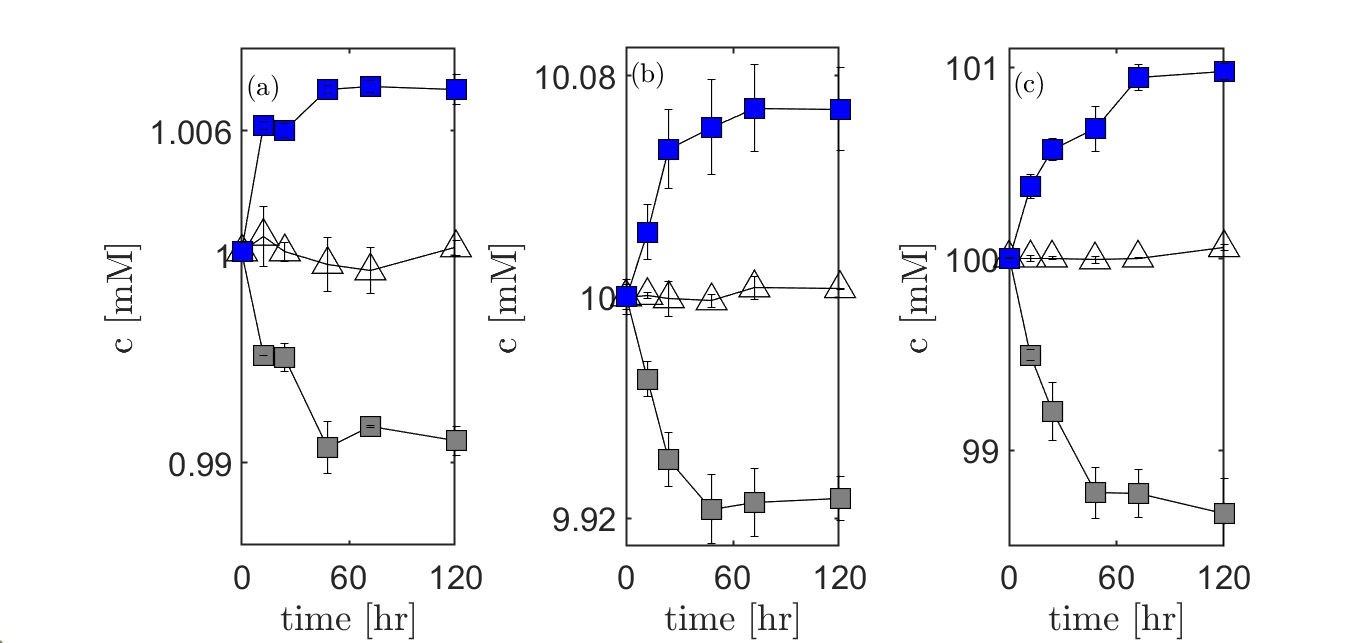}
      \caption{Temporal evolution of diamagnetic ZnCl$_{2}$ salt concentration (squares) in polydisperse silica gel induced by a NdFeB permanent magnet ($\mathbf{B}$ = 0.74 T, \BgradB$|_{max}$ = 100 T$^{2}$/m) for initial concentrations of (a) 1 mM (b) 10 mM, and (c) 100 mM. Markers are presented as follows: blue is far section, gray is near section, and triangle denotes the control experiments with $\mathbf{B}$ = 0. }
    \label{fig:ZnResults}
\end{figure}

Next, we investigate solutions containing the single diamagnetic ZnCl\(_2\) ion. Fig.~\ref{fig:ZnResults}(a-c) shows the temporal evolution of the diamagnetic ZnCl\(_2\) ion concentration in the near and far sections of the measuring cell. The experiments reveal that the diamagnetic metal ions migrate towards regions of weaker magnetic fields, or away from regions with maximum magnetic field gradients. This migration is observed as an enrichment of the metal solution in the far section of the silica gel. After 72 hours of magnetic field exposure, a moderate concentration depletion of 0.6 - 1.2\% is observed in the near slice, within high-field regions. Notably, the amount of metal ion depletion does not follow a monotonic trend with respect to the initial ZnCl\(_2\) concentration. Additionally, no concentration changes were observed in control experiments conducted without magnetic field exposure (\(|\mathbf{B} \cdot \nabla \mathbf{B}| = 0\)), confirming that the observed concentration changes are attributed to the external magnetic field.

\subsection{Magnetophoresis of single metal ions in monodisperse silica gel} 

\begin{figure}[hthp]
    \centering
    \begin{subfigure}{1\textwidth}
        \centering
        \includegraphics[width=\textwidth]{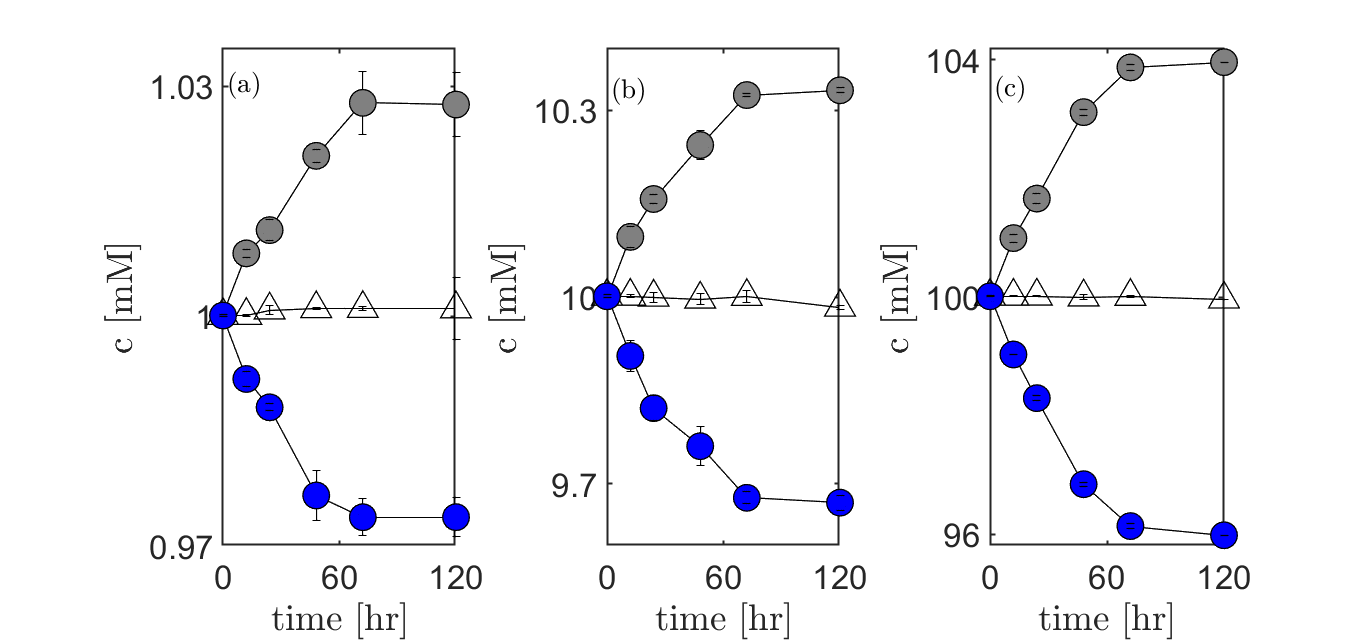}
        \label{fig:sub1}
    \end{subfigure}        
    \begin{subfigure}{1\textwidth}
         \centering
        \includegraphics[width=\textwidth]{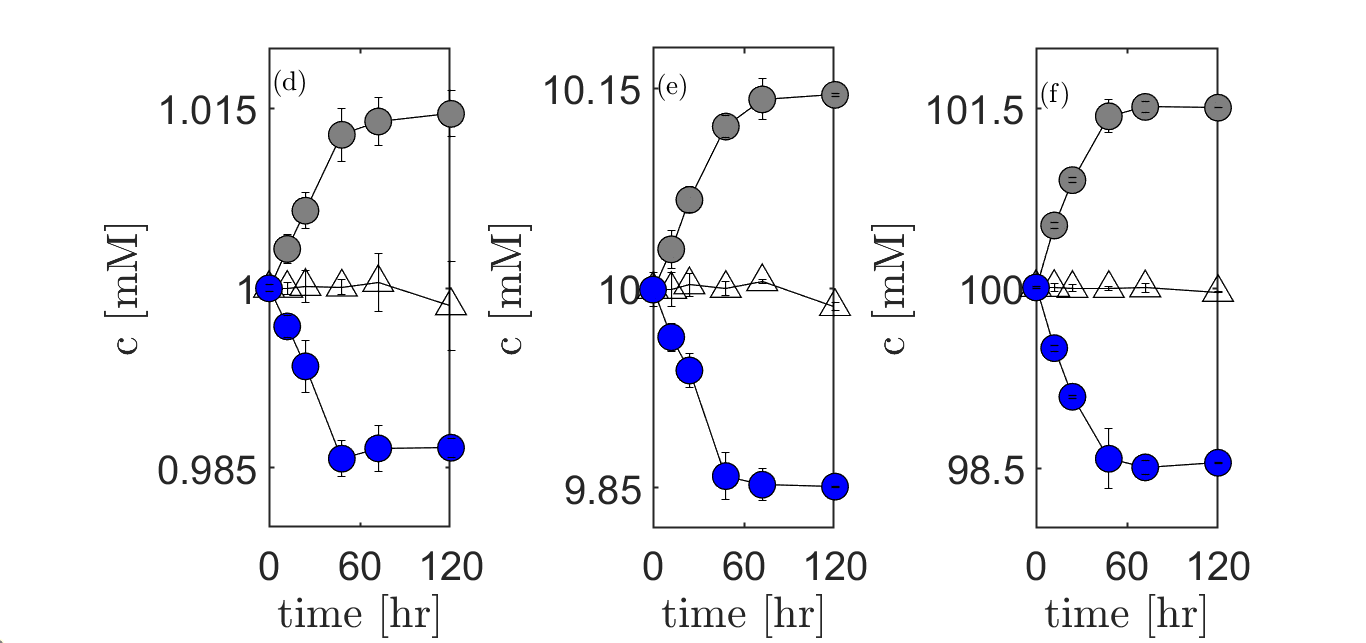 }
        \label{fig:sub1}
    \end{subfigure}
 \caption{Temporal evolution of concentration for paramagnetic MnCl$_{2}$ salt in 500 $\mu$m (a-c) and 63 $\mu$m (d-f) silica gels induced by a NdFeB permanent magnet ($\mathbf{B}$ = 0.74 T, \BgradB$|_{max}$ = 100 T$^{2}$/m). Here blue, gray circles and triangles correspond to far section, near section, and the control experiments.}
    \label{fig:PS_MnResults}
\end{figure}

Next, we investigate the impact of porous media particle size on the magnetophoresis of metal ions. Fig.~\ref{fig:PS_MnResults} shows the temporal evolution of MnCl\(_2\) concentration in silica gels with nominal particle sizes of 63 $\mu$m and 500 $\mu$m, for initial concentrations of 1mM, 10mM, and 100mM. Similar to experiments conducted with silica gels of broader particle size distributions, the paramagnetic MnCl\(_2\) ion migrates towards regions of high magnetic fields. For the 500 $\mu$m silica gel, concentration enrichment of approximately 3-4\% was observed, slightly higher than the enrichment level recorded for the polydisperse silica gel. In contrast, the 63 $\mu$m silica gel exhibited a lower concentration enrichment of 1.3–1.5\%, which is less than the enrichment observed in both the polydisperse silica gel and the 500 $\mu$m silica gel particles.
Control experiments without magnetic field exposure, conducted for MnCl\(_2\) using both silica sizes, showed no concentration changes, confirming that the metal ion transport is due to the influence of the external magnetic field gradient. These results suggest that the size of the silica particles plays a role in the magnetophoresis of paramagnetic metal ions, with smaller particle sizes leading to less enrichment of the metal ion near the magnet surface.

\begin{figure}[hthp]
    \centering
    \begin{subfigure}{1\textwidth}
        \centering
        \includegraphics[width=\textwidth]{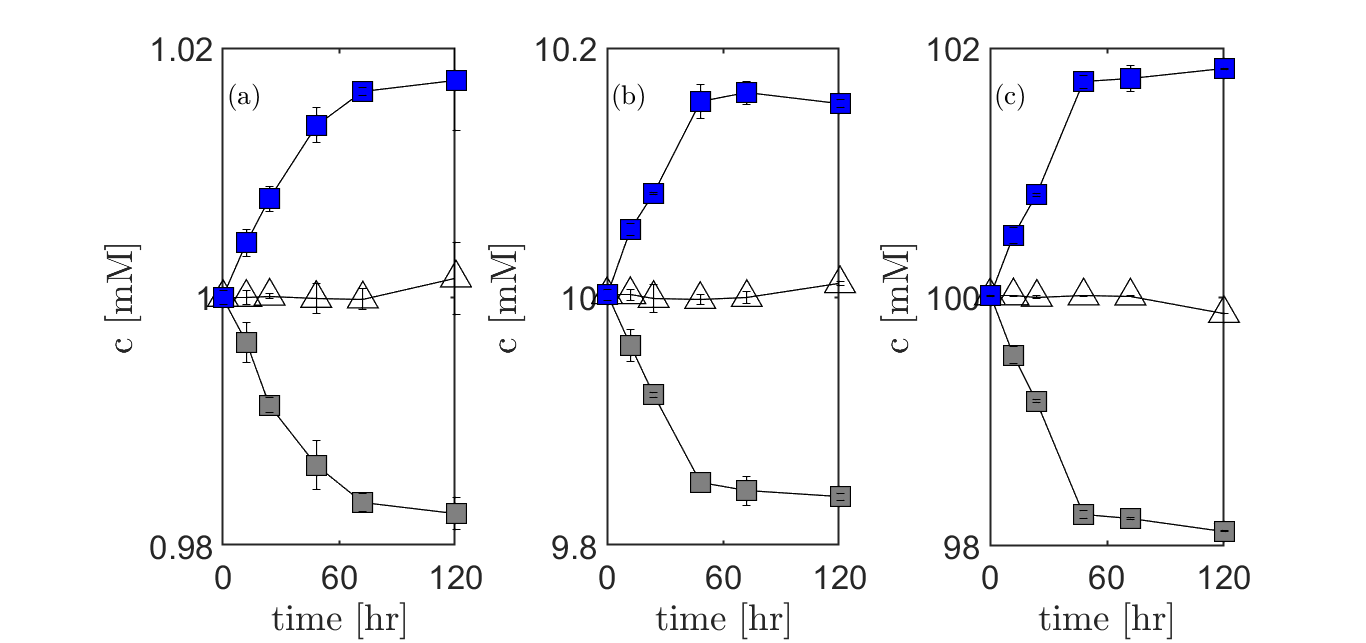}
        \label{fig:sub1}
    \end{subfigure}
        \begin{subfigure}{1\textwidth}
         \centering
        \includegraphics[width=\textwidth]{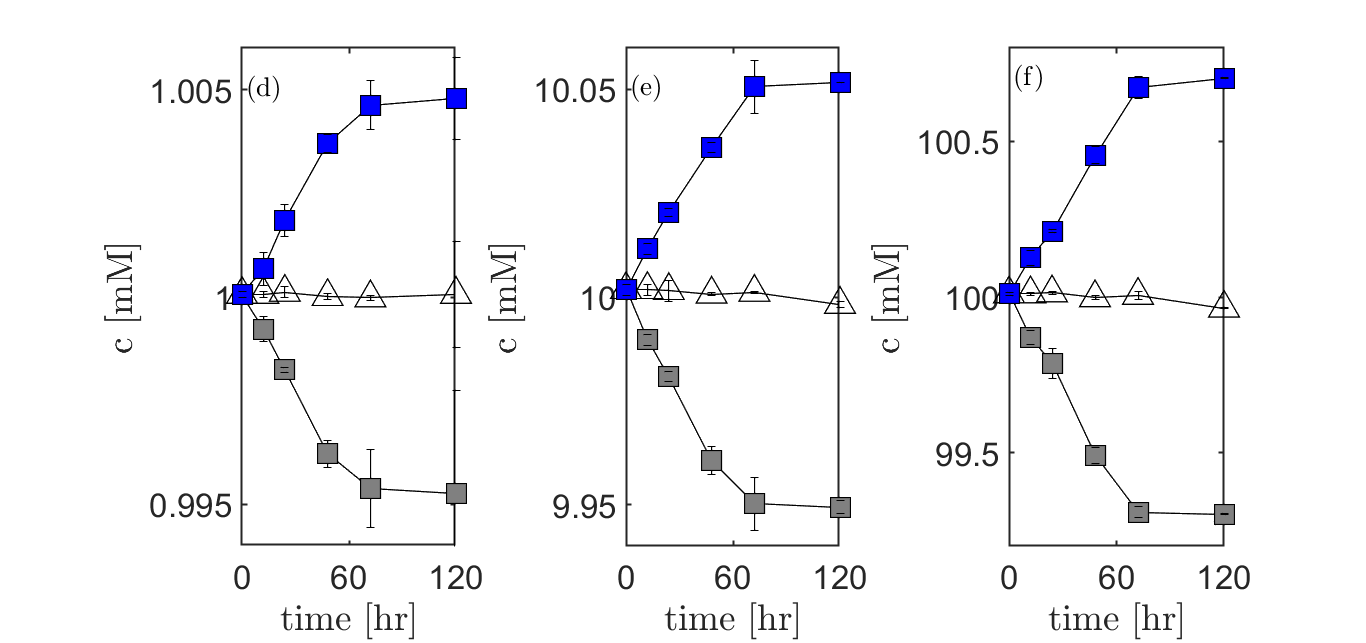}
        \label{fig:sub1}
    \end{subfigure}
\caption{Temporal evolution of concentration for paramagnetic ZnCl$_{2}$ salt in 500 $\mu$m (a-c), and 63 $\mu$m (d-f) silica gel induced by a NdFeB permanent magnet ($\mathbf{B}$ = 0.74 T, \BgradB$|_{max}$ = 100 T$^{2}$/m).  Here blue, gray squares and triangles correspond to far section, near section, and the control experiments.}
    \label{fig:PS_ZnResults}
\end{figure}

In addition, Fig.~\ref{fig:PS_ZnResults} illustrates the temporal evolution of diamagnetic ZnCl\(_2\) concentration in silica gels with 63 $\mu$m and 500 $\mu$m particle sizes, using an initial concentration of 1mM, 10mM, and 100 mM. Similar to the experiments conducted with polydisperse silica gels, the ZnCl\(_2\) ion is depleted in regions of high magnetic fields and accumulates (or is enriched) in the far section of the silica gel slice. A moderate depletion of 1.6-1.8\% is observed for ZnCl\(_2\) in the 500 $\mu$m silica gel, slightly exceeding the depletion levels recorded for ZnCl\(_2\) in the polydisperse silica gel and the 63 $\mu$m silica gel, which exhibited depletion levels of approximately 0.5-0.7\%. Furthermore, the control experiments showed no concentration variation over time, reinforcing that the observed changes in the temporal evolution of diamagnetic ZnCl\(_2\) concentration are directly linked to exposure to the external magnetic field.

\begin{figure}[hthp]
    \centering
       \includegraphics[width=\textwidth]{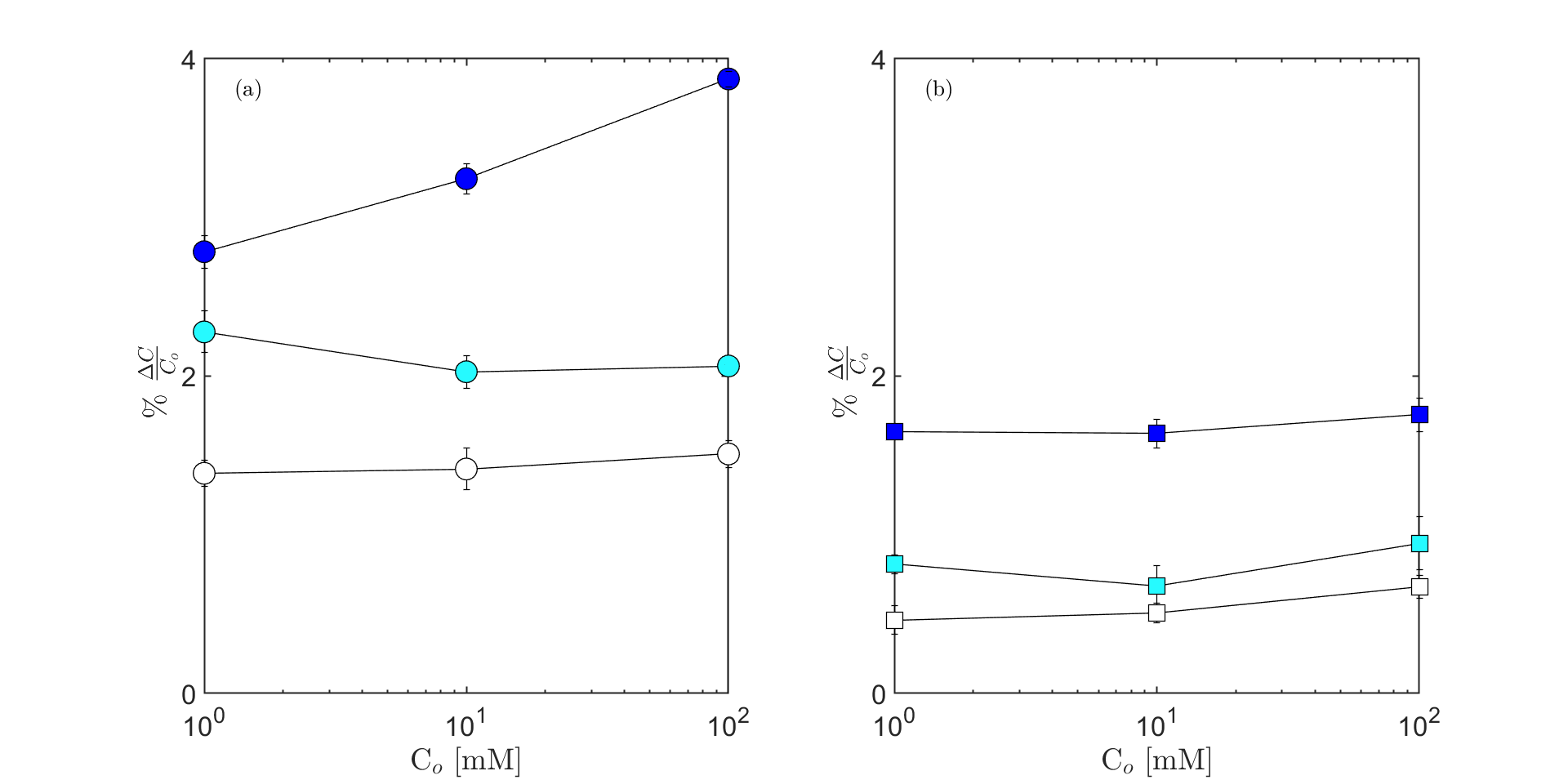}
\caption{Normalized concentration changes for MnCl$_2$ in the near section and ZnCl$_2$ in the far section after 72 hours of exposure to the magnetic field, as a function of initial ion concentration for various silica gel particle sizes. Markers are presented as follows: white denotes 63 $\mu$m silica particle, cyan represents polydisperse silica gel, and blue corresponds to 500 $\mu$m silica gel particles.} 
    \label{fig:Mn_Zn_summary}
\end{figure}

Fig.~\ref{fig:Mn_Zn_summary} provides a summary of the normalized concentration variation for both paramagnetic and diamagnetic metal ions across different porous media and the tested range of initial concentrations. For paramagnetic MnCl$_2$ (Fig.~\ref{fig:Mn_Zn_summary}(a)), the highest concentration enrichment is observed in the porous media with the largest particle size (500 $\mu$m), and the enrichment levels increase with higher initial concentrations of the ion. In contrast, for porous media with smaller particle sizes (63 $\mu$m) or polydisperse particles, the overall concentration enrichment is comparatively lower and shows minimal dependence on the initial ion concentration. For diamagnetic ZnCl$_2$ (Fig.~\ref{fig:Mn_Zn_summary}(b)), the depletion near the magnet surface follows a similar trend, with the greatest depletion occurring in the porous media with the largest particle size. However, unlike MnCl$_2$, the depletion levels of ZnCl$_2$ remain relatively constant throughout the range of initial ion concentrations tested, regardless of the particle size of the porous medium. \HM{This result is consistent with previous observations of magnetophoretic motion of superparamagnetic and weakly paramagnetic nanoparticles in dilute suspensions, where the initial concentration did not influence the time-dependent variation of the normalized particle concentration under an applied magnetic field~\cite{leong2015magnetophoresis,Rassolov25}.}

\par

\subsection{Magnetophoresis of solutions containing binary metal ions}
\begin{figure}[hthp]
    \centering
    \begin{subfigure}{0.75\textwidth}
        \centering
        \includegraphics[width=\textwidth]{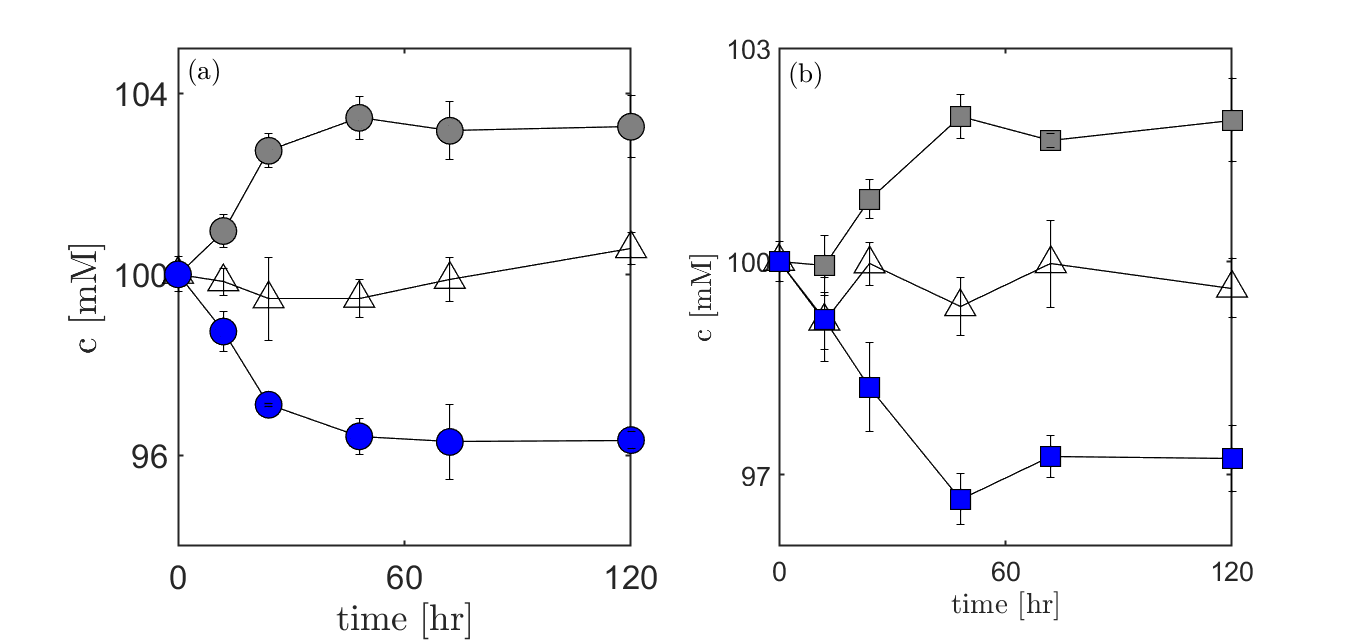}
        \label{fig:sub1}
    \end{subfigure}
    \begin{subfigure}{0.75\textwidth}
        \centering
        \includegraphics[width=\textwidth]{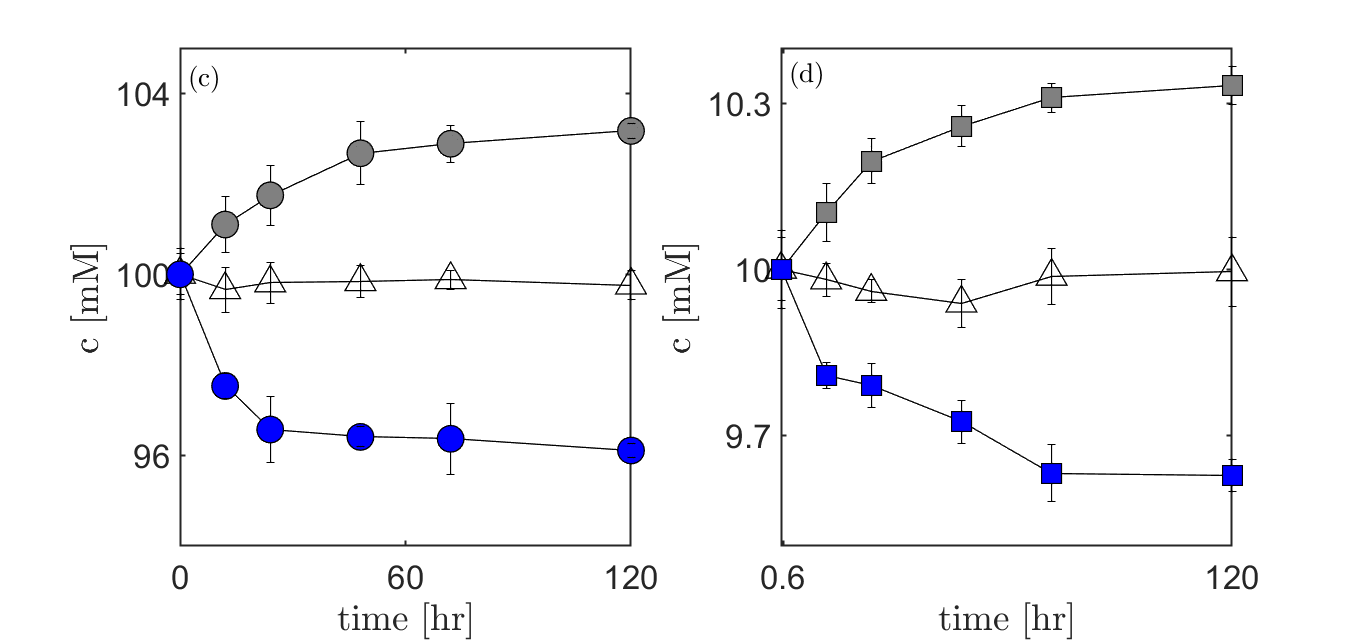}
    \end{subfigure}
    \begin{subfigure}{0.75\textwidth}
        \centering
        \includegraphics[width=\textwidth]{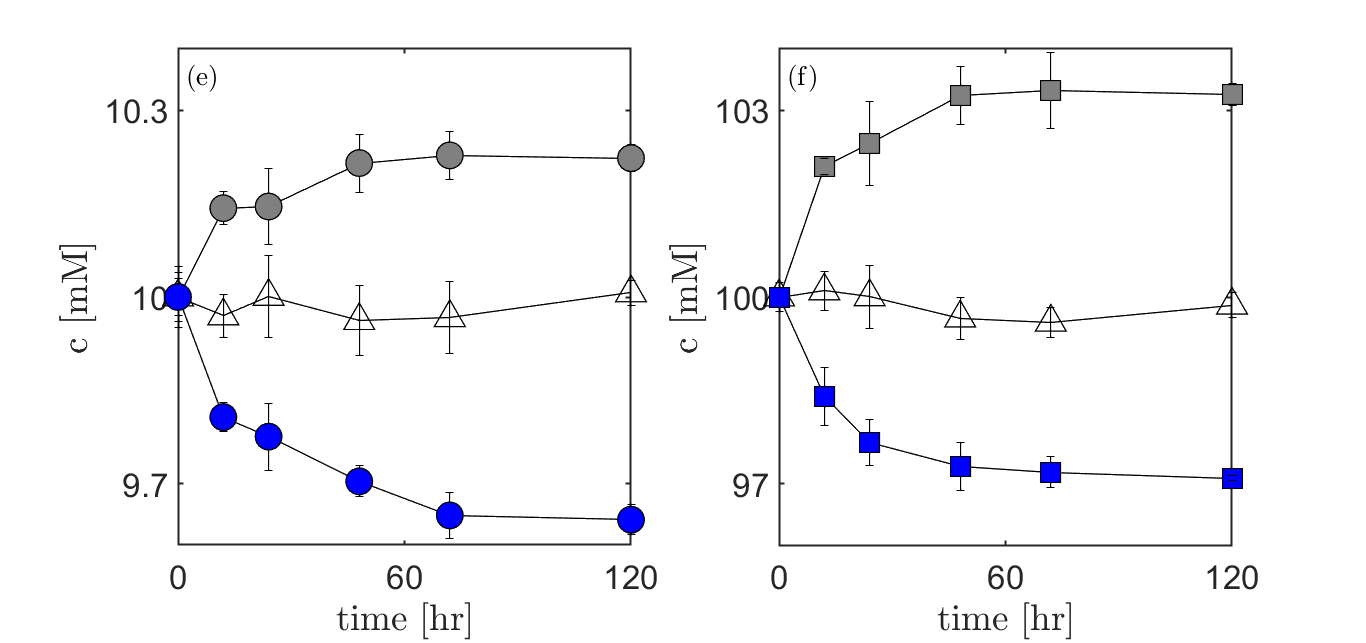}
        \label{fig:sub3}
    \end{subfigure}
    \caption{Temporal evolution of concentration in a MnCl$_2$  (circle) and ZnCl$_2$ (square) binary mixtures induced by a NdFeB permanent magnet ($\mathbf{B}$ = 0.74 T, \BgradB$|_{max}$ = 100 T$^{2}$/m). Mixture conditions: (a,b) equimolar 100 mM MnCl$_2$:100 mM ZnCl$_2$, (c,d) 100 mM MnCl$_2$:10 mM ZnCl$_2$, and (e,f) 10 mM MnCl$_2$:100 mM ZnCl$_2$. Markers are presented as follows: blue denotes far section, gray represents the near section, and triangles indicate control experiments with $\mathbf{B}$ = 0.}
    \label{fig:MixResults}
\end{figure}

Following the studies on single metal ions, the magnetophoresis behavior was investigated in solutions containing both metal ions (i.e., binary mixtures). Binary mixtures were prepared with MnCl\(_2\):ZnCl\(_2\) concentration ratios of 1:1, 1:10, and 10:1. For these experiments, 500 \(\mu\)m silica gel was used as the porous medium, as it demonstrated the highest levels of magnetophoresis for both metal ions in single-ion studies. Fig.~(\ref{fig:MixResults}) shows the temporal evolution of each metal ions in the mixture for the near and far section of the silica gel. We start describing the results of the binary solution with equimolar concentration of MnCl$_2$ and ZnCl$_2$ (100mM). Figure~\ref{fig:MixResults}(a) illustrates the temporal evolution of MnCl\(_2\) concentration under the influence of an external magnetic field. Consistent with individual metal ion experiments, the paramagnetic MnCl\(_2\) ions migrate towards the magnet surface, resulting in enrichment in the near section of the silica gel. Surprisingly, as shown in Figure~\ref{fig:MixResults}(b), the diamagnetic ZnCl\(_2\) ions also exhibit enrichment near high-field regions, contrary to their behavior in single-ion experiments. Specifically, in the equimolar mixture, MnCl\(_2\) achieves a 3.5\% enrichment relative to its initial concentration, while ZnCl\(_2\) shows a 2\% enrichment after 72 hours. These findings highlight the critical role of interactions between the two metal ions in determining their magnetophoresis behavior in mixtures. \par 
To further investigate the impact of ion-ion interactions, additional experiments were carried out using solutions with 100:10 and 10:100 molar ratios of MnCl\(_2\):ZnCl\(_2\). Figures~\ref{fig:MixResults}(c,f) present the results for these mixtures. As shown in Fig.~\ref{fig:MixResults}(c,d), for the 100:10 MnCl\(_2\):ZnCl\(_2\) mixture, MnCl\(_2\) achieved a 2.8\% enrichment, while ZnCl\(_2\) concentration in the near section increased by 3.1\% after 72 hours. Similarly, in the 10:100 MnCl\(_2\):ZnCl\(_2\) mixture (Fig.~\ref{fig:MixResults}(e,f)), MnCl\(_2\) reached an enrichment level of 2.2\%, while ZnCl\(_2\) achieved an enrichment of 3.2\% after 72 hours. Control experiments performed without a magnetic field gradient showed no concentration changes for either ion, confirming that the observed results are solely attributable to the magnetic field and not to adsorption or evaporation effects.\par  



\section{Discussion}
Our experimental results confirm the occurrence of magnetophoresis for both weakly paramagnetic (MnCl$_{2}$) and diamagnetic (ZnCl$_{2}$) metal ions, irrespective of the porous medium used or whether the solutions consist of single or binary components. To gain a deeper insight into the underlying physics governing the experimental observations, we analyze the forces involved in the process. We begin by discussing the experimental results for the single-metal ion systems, followed by our interpretation of the trends and results for binary mixtures.\par 

The paramagnetic and diamagnetic transition metal salts, when dissolved in water, dissociate into ions and will be hydrated with water molecules. The radius of a hydrated ion surrounded by water molecules could be approximately close to 3 $\r{A}$. Beyond this estimation, it is also possible that a partial second hydration shell, including the chloride anions could be involved, for this small cation, leading to a $b \approx$ 6 $\r{A}$\cite{marcus1988ionic}. Using the maximum magnetic field gradients measured in the experimental cell and shown in Fig.~\ref{2DBgradient}(c), we can compute the magnetic \Peclet numbers to be Pe$_{m}|_{max} \approx \mathcal{O}({10^{-10}})$ and $\mathcal{O}({10^{-7}})$ for ZnCl$_{2}$ and MnCl$_{2}$, respectively. \HM{As noted before, the Brownian force was estimated to be in the order of $k_BT/b$. Alternatively, one can estimate the root-mean-square Brownian force magnitude by using fluctuation-dissipation theorem as: $F_{\rm RMS} \approx \sqrt{{2 k_B T (6\pi \eta b)}/{\Delta t}}$~\cite{kubo1966fluctuation}. Over a short time interval of $10^{-9}$ sec (that corresponds to a diffusion in order of metal ion hydration size), the force can be estimated to be about $\mathcal{O}(10^{-11})$N, which is in the same order of magnitude as of $k_BT/b$. Prior studies have also used the ratio of the magnetic energy to thermal diffusion energy ($k_BT$) as a method to evaluate the strength of magnetic forces to Brownian diffusive forces~\cite{rodrigues2017}. Our calculations indicate that the ratio of these two energies is much smaller than unity.} Therefore, it is unlikely that the magnetic force acting on a single metal ion is sufficient to overcome diffusive forces and induce magnetophoresis in the experiments reported in this study. \HM{Note that this conclusion is consistent with prior published literature that the magnetic force (or energy) on individual metal ions is not strong enough to overcome the thermal fluctuations assoictaed with Brownian motion~\cite{rodrigues2017,lei2017}.} \\

Magnetophoresis is fundamentally achievable only when the magnetic \Peclet number Pe$_m \geq$ 1. Our hypothesis posits that under the experimental conditions of this study, magnetophoresis occurs only when the effective size of the magnetic metal ion, $b$, is sufficiently increased, typically through the formation of magnetic field-induced clusters or aggregates. This field-induced clustering phenomenon amplifies the magnetic force acting on the metal ion, thereby enhancing the likelihood of magnetophoresis. From our analysis, the minimum estimated cluster sizes required for observable magnetophoresis (Pe$_m$ = 1) are approximately $b$ $\approx$ 0.05 $\mu$m and $b$ $\approx$ 0.22 $\mu$m for MnCl$_{2}$ and ZnCl$_{2}$, respectively. While our analysis establishes a lower threshold for the cluster sizes required for magnetophoresis, it does not quantitatively account for the dynamic evolution of cluster size as a function of time and metal ion concentration and its variation within the domain. In part II of this study, we will provide a comprehensive quantitative analysis using multi-physics numerical simulations to explore how metal ion cluster size evolves as a function of magnetic susceptibility, initial ion concentration, and time. This will offer deeper insight into the mechanisms underlying magnetophoresis in porous media containing metal ions.\par 

The concept of field-induced cluster formation is not novel and has been invoked in prior magnetophoresis studies. For instance, Fujiwara and colleagues \cite{fujiwara2001,chie2003,fujiwara2006} observed magnetophoretic velocities significantly exceeding predictions based on a single metal ion force balance when paramagnetic ions were exposed to magnetic field gradient generated by a superconducting magnet. Although theoretical calculations suggested no detectable movement for their metal ion, the experimental data demonstrated a pronounced magnetophoretic velocity, implicating cluster formation as the driving mechanism. Franczak et al.~\cite{franczak2016} proposed field-induced cluster formation to explain similar experimental observation for other types of metal ions in porous media. On the other hand, several studies have investigated the conditions required for field-induced cluster formation under a uniform magnetic field for both colloidal and non-colloidal particles\cite{faraudo2013,faraudo2016,heinrich2015}. Faraudo and colleagues \cite{faraudo2013,faraudo2016} reported that superparamagnetic particles form aggregates when exposed to an external magnetic field, primarily driven by dipole-dipole interactions. The significance of this field-induced aggregation, however, is contingent on the relative strength of the magnetic interactions compared to the thermal energy. The relative strength of these two energies can be expressed through a dimensionless magnetic coupling parameter, $\Gamma$, which is defined as \cite{faraudo2013,heinrich2015,faraudo2016,andreu2011,leong2020unified}
\begin{equation}
\label{gamma}
    \Gamma = \frac{\mu_{0} m_{s}^{2}}{2\pi b^3 k_{B} T}.
\end{equation}
where $m_s$ is the magnetization of the magnetic particles. When $\Gamma >$ 1, magnetic dipole-dipole interactions dominate, whereas in the opposite limit, $\Gamma <$ 1, thermal dissociation prevails. However, experimental results have shown that field-induced aggregate formation does not always occur, even in cases where $\Gamma > 1$~\cite{faraudo2013}. 
It has been suggested that the rate of aggregate formation under a uniform magnetic field must surpass the rate of particle dissociation, which can be quantified using a new dimensionless aggregation parameter as:\cite{faraudo2013,heinrich2015,faraudo2016,andreu2011,leong2020unified}
\begin{equation}
\label{gamma}
    N^* = \sqrt{\phi_{o} e^{\Gamma -1}}
\end{equation}
Here, $\phi_{o}$ represents the volume fraction of magnetic particles in the solution. Therefore, for cluster formation to occur, the conditions $\Gamma > 1$ and $N^* > 1$ must be satisfied. Using the magnetic flux density and the concentrations of the metal ions, we can compute these dimensionless numbers $\Gamma|_{max} \approx \mathcal{O}({10^{-8}})$ and $N^{*}|_{max} \approx \mathcal{O}({10^{-2}})$ for MnCl$_{2}$ metal ions. Based on these dimensionless parameters, field-induced metal ion cluster formation is unlikely under our experimental conditions and even experiments of Fujiwara et al.~\cite{fujiwara2004,fujiwara2006} and Franczak et al \cite{franczak2016}. Note that the above dimensionless numbers do not account for a non-uniform magnetic field, which is expected to generate a magnetophoretic force, thereby enhancing ion transport and potentially facilitating field-induced cluster formation. \HM{As noted above, for micro- and nano-particles, field-induced clustering can be directly visualized using optical microscopy~\cite{faraudo2016,andreu2011}. However, this approach is not applicable to metal ions, whose characteristic length scales lie far below the diffraction limit of light. At the atomic level, the most direct probes of liquid structure via X-ray,  neutron diffraction, and/or dynamic light scattering, are powerful tools for characterizing correlations in complex fluids. Yet, in the case of dilute metal ions, these techniques are fundamentally limited: the scattering signal arising from ion-ion correlations is vanishingly small compared to the dominant contributions from water-water and water-ion interactions~\cite{hassan2008computer}. Consequently, such methods cannot unambiguously resolve whether metal ions form clusters under applied fields. Most recently, Bian et al.~\cite{bian2011ion} employed a vibrational energy transfer approach to demonstrate the formation of ion clusters in aqueous solutions driven by electrostatic interactions. They reportedly observed ion clustering in aqueous solutions of medium to high concentrations, and diluted solutions produced fewer and smaller clusters. This technique represents a promising avenue for probing ion–ion correlations or clustering through experimental means; however, its applicability to the investigation of metal ion clustering under the influence of external magnetic fields has yet to be systematically assessed.}\\


Furthermore, our results in Fig.~\ref{fig:Mn_Zn_summary} show that increasing silica particle size strengthens magnetophoresis and enhances ion transport through porous media for both paramagnetic MnCl$_{2}$ and diamagnetic ZnCl$_{2}$. Our initial hypothesis is that an increase in silica particle size alters the medium's effective porosity, which in turn may impact metal ion diffusion and modify the resulting diffusive forces. This change directly impacts the magnetic \Peclet number. The magnetic \Peclet number $Pe_m \propto  1/D_e$\cite{rassolov2024magnetophoresis}, where $D_e$ denotes the effective diffusion coefficients of metal ions in the porous media. The pore spaces between silica gel particles are interconnected and vary in diameter. Assuming an average pore diameter, the effective diffusion coefficient for randomly distributed pores can be approximated as~\cite{welty2014,millington1961permeability}: $D_e = \varepsilon^{4/3} D_{f}$\cite{welty2014}. Here $D_f$ is the diffusion coefficient in free space, and $\varepsilon$ is the experimentally determined porosity. The porosity can be estimated as \cite{welty2014}:

\begin{equation}
\label{porosity}
    \varepsilon = \frac{V_{\text{ solution}}}{V_\text{solution}+V_\text{silica}}.
\end{equation}
Here, $V_\text{solution}$ represents the volume of the metal ion solution used to prepare the silica gel bed, while $V_\text{silica}$ denotes the volume occupied by the silica beads. Our measurements indicate an effective porosity value are 0.71 for 500 $\mu$m silica particles and 0.74 for 63 $\mu$m silica particles. The similarity in porosity values indicates that porosity alone cannot fully account for the observed influence of porous media particle size on the magnetophoresis of metal ions. This suggests that additional factors likely contributed to the experimental results shown in Fig.~(\ref{fig:Mn_Zn_summary}).  \par 

In the above analysis, we assumed that the porous medium could be treated as an effective continuum domain, influencing magnetophoresis primarily through the diffusion of metal ions, while neglecting any direct effect on the drag force. However, in a porous medium, the magnetic force acting on an ion in the embedded fluid is counteracted by a drag force $F_d$, which arises not only from the viscous forces in the fluid but also from interactions between the ion clusters and the surrounding porous matrix wall. Brinkman demonstrated that this drag force represents a modified form of the Stokes' drag equation \cite{brinkman1949}, accounting for the additional resistance imposed by the porous structure as:
\begin{equation}
\label{dragforce}
    F_d = 6 \pi \eta b v_{mig} \left [ 1+\lambda b+\frac{\lambda^{2} b^{2}}{3} \right ],
\end{equation} 
where $v_{mig}$ denotes the magnetophoresis velocity of the ion, $\eta$ represents the viscosity of the fluid environment, and $\lambda$ is a parameter related to the permeability of the porous medium $\lambda^2 \propto \left (\frac{1}{k} \right ) $\cite{brinkman1949}. Here $k$ is the permeability of the porous medium, and it may be estimated using the Carman-Kozeny relation as\cite{schulz2019,rehman2024}. 
\begin{equation}
    k =  \frac{d_{p}^{2}}{180} \frac{\varepsilon^3}{(1-\varepsilon)^{2}}.
    \label{permeability}
\end{equation}
Here, $d_p$ represents the average size of the porous medium particles. In the limit of high permeability (i.e., low porous medium particle density, \( k \to \infty \), \( \lambda \to 0 \)), the drag force acting on the ion reduces to Stokes' drag. Based on the above relations, the drag force experienced by a solute in a porous medium scales inversely with the porous media particle size, i.e., $F_d \sim 1/d_p$. Considering a steady-state force balance on a metal ion cluster subjected to a magnetic field, the magnetic driving force and hydrodynamic drag satisfy $F_m - F_d = 0$. Under this balance, the magnetophoretic velocity scales as $v_{mig} \sim d_p$, indicating that larger porous media particle diameters reduce hydrodynamic resistance. Consequently, increasing the characteristic size of silica particles within the medium is expected to enhance their magnetophoretic velocity. This trend aligns with our experimental data shown in Fig.~\ref{fig:Mn_Zn_summary}, which demonstrates stronger magnetophoresis for silica gel with 500 $\mu$m particles compared to 63 $\mu$m silica particles.\par 

Although the above analysis qualitatively explains the experimental trends, it highlights the necessity of incorporating the permeability of the porous medium into the force balance on metal ions. A simple Stokes' drag model assumes a homogeneous, unbounded fluid environment and neglects the influence of the porous structure, which can significantly alter momentum transfer. In our previous model, we used simple Stokes' drag to compute the magnetophoresis velocity of metal ions in the porous media\cite{rassolov2024magnetophoresis}. Given that metal ions migrate through a confined and heterogeneous porous network, their motion is subjected to additional resistance beyond that predicted by Stokes’ drag alone. The Brinkman equation extends Darcy’s law by incorporating viscous effects within the porous medium, effectively bridging the gap between Stokes' drag in free fluid and Darcy’s law in highly permeable media. This approach provides a more comprehensive description of the momentum and mass transfer by accounting for the local velocity gradients and the effective viscosity within the porous network. Therefore, to accurately capture the experimental observations, the force balance must incorporate a porous media model, such as the Brinkman equation, rather than relying solely on Stokes' drag. In the second part of this study, we will systematically compare the predictions of Stokes' drag, with those of the Brinkman equation, which accounts for the influence of porous media permeability on momentum transfer. This comparison will allow us to assess the validity and limitations of each approach in describing the magnetophoretic transport of metal ions in porous media.\par 

Finally, we examine the results for binary mixtures of metal ions, which exhibit intriguing and unexpected behaviors. Notably, in these mixtures, both paramagnetic and diamagnetic ions exhibit migration toward regions of the highest magnetic field gradients (i.e., the surface of the permanent magnet). This behavior is particularly surprising for diamagnetic ZnCl$_2$, which is expected to move away from regions of high magnetic field. In addition, the experimental results indicate that the concentration variations of paramagnetic MnCl$_2$ in binary mixtures are less pronounced than those in experiments involving single metal ions, and are even weakened when the initial concentration of ZnCl$_2$ in the mixture increases. These observations suggest that strong ionic interactions between metal ions in binary mixtures may lead to cluster formation containing both paramagnetic and diamagnetic species.\\ 

\HM{It is worth noting that the paramagnetic Mn$^{+2}$ and diamagnetic Zn$^{+2}$ metal ions have similar charges. These metal ions in solution are expected to interact with each other through repulsive electrostatic forces~\cite{neu1999,trizac1999,sader1999,trizac2000}. Therefore, formation of ion clusters containing both metal ions seem non-intuitive at first glance. However, several studies have reported electrostatic attraction of similarly charged species in solutions \cite{naji2011exotic,naji2004attraction,naji2005electrostatic,ghodrat2015strong,kanduvc2010counterion,linse2000electrostatic,moreira2002simulations}, implying that the coexistence of these paramagnetic and diamagnetic ions within the same cluster is possible. Previous studies have shown that the strength of electrostatic interactions between charged species can be quantified by the dimensionless electrostatic coupling parameter, $\Xi = l_B/\mu$. The coupling parameter is defined as the ratio of the Bjerrum length to the Gouy–Chapman length. The Bjerrum length, $l_B$, which specifies the distance at which two
charges interact with an energy equal to $k_BT$, and is expressed as: \cite{kanduvc2010counterion,naji2004attraction,naji2005electrostatic}
\begin{equation}
    l_B = \frac{q^2 e^2}{4 \pi \epsilon \epsilon_0 k_B T},
\end{equation}
where $q$ is the counterion charge valency (Cl$^{-}$), $e$ is the elementary charge ($1.602 \times 10^{-19}$ C) \cite{jeckelmann2019elementary}, $\epsilon$ is the dielectric constant at room temperature (80)\cite{naji2005electrostatic}, and $\epsilon_0$ is the permittivity of vacuum ($8.854 \times 10^{-12}$ F/m)\cite{komarov2005permittivity}.\\

On the other hand, the Gouy-chapman length, $\mu$, is defined as: \cite{kanduvc2010counterion,naji2004attraction,naji2005electrostatic,naji2011exotic}
\begin{equation}
    \mu = \frac{1}{2 \pi q l_B \sigma_s},
\end{equation}
where $\sigma_s$ is the surface charge density\cite{naji2004attraction}. The surface charge density is estimated as $\sigma_s = Z/4\pi b^{2}$; where $Z$ is the surface charge valency (Mn$^{2+}$/Zn$^{2+}$) \cite{naji2004attraction}. The Gouy-chapman length measures the distance at which the thermal energy equals the electrostatic energy between the surrounding couterions and the elementary charge. Based on the above analysis, it has been predicted that for $\Xi > 1$ values the like-charged species experience electrostatic attraction~\cite{naji2004attraction}. Considering the numerical values of the charge valency of the two species and their hydration radius, the estimated coupling parameter for the MnCl$_{2}$-ZnCl$_{2}$ binary mixture is $\Xi \approx 8.9$. This indicates that the Mn$^{+2}$ and Zn$^{+2}$ like charged ions may attract each other and support the above hypothesis that they may form a cluster that contains both of these metal ions. }\\

\HM{The above analysis is relevant when species are considered to have flat surfaces with no curvature. Since the elementary charge has a curved surface, such as spherical geometries, the radius of curvature $\hat{R}$ introduces a new length scale that may influence counterion behavior. When curvature $\hat{R}$ is much larger than the Gouy-Chapman length, $\mu$, the system behaves similarly to the planar case. Deviations from planar behavior arise when $\hat{R} \approx \mu$ or smaller. It has been shown that a useful parameter to describe curvature effects is the Manning parameter \cite{naji2004attraction} defined as: $ \xi = {\hat{R}}/{\mu}$.
For $\xi > 1$, counterions condense around the elementary charge, leading to electrostatic attraction. In contrast, when $\xi < 1$, counterions de-condense and diffuse into the bulk fluid, and no attraction is expected \cite{naji2004attraction}. Thus, in addition to coupling parameter $\Xi$, for electrostatic attraction between like charges in curved surfaces, the Manning parameter $\xi$ must be greater than unity. From the charge valency of the two species and their hydration radius, the estimated Manning parameter for the MnCl$_{2}$-ZnCl$_{2}$ binary mixture is $\xi \approx 2.98$. This suggests that manganese and zinc ions at experimental conditions, are likely to experience electrostatic attractions, favoring the coexistence of diamagnetic and paramagnetic ions within the same cluster.}\\

We hypothesize that such clustering influences magnetophoresis by modifying the effective magnetic susceptibility of the transported species. Specifically, the magnetic susceptibility of a cluster can be expressed as: $\chi_{c} = \sum_{i=1}^{2} {y_i \chi_i}$. Here $y_i$ and $\chi_i$ denote the mole fraction and magnetic susceptibility of each metal ions in the binary mixture, respectively. Given the significantly higher magnetic susceptibility of MnCl$_2$, we propose that MnCl$_2$ rich clusters dominate the magnetophoretic motion, effectively driving the entire cluster toward regions of higher magnetic field gradients. This hypothesis implies that both metal ions could exhibit magnetophoretic motion towards high magnetic field gradients despite ZnCl$_2$ being diamagnetic. Furthermore, as the ZnCl$_2$ concentration increases, the cluster's overall magnetic susceptibility decreases, weakening the net magnetic force. This is consistent with our experimental observations, which show reduced MnCl$_2$ enrichment near the magnet surface at higher ZnCl$_2$ concentrations.\par

\section{Summary and conclusions}

In this study, we investigated the magnetophoresis of paramagnetic and diamagnetic metal ions within a porous medium exposed to the non-uniform magnetic field of a permanent magnet. The key findings from our experiments are summarized below:\par 

For experiments involving single metal ions, paramagnetic metal ions based on MnCl$_2$ were observed to migrate toward regions of the highest magnetic field gradients, achieving moderate enrichment levels of approximately 2–4\% across initial concentrations of 1-100 mM. Conversely, diamagnetic ZnCl$_2$ ions migrated away from high field regions, leading to measurable depletion (0.6-1.8\%) near the magnet surface. Our force balance analysis indicates that these concentration changes are likely facilitated by field-induced cluster formation, where ion clusters experience enhanced magnetic forces. Furthermore, increasing the porous media particle size enhanced the extent of magnetophoresis for both paramagnetic and diamagnetic ions. This behavior can be attributed to the reduction in drag forces experienced by the ions in a porous media with larger permeability, thereby allowing greater mobility in response to the magnetic field.\par 

Binary mixtures of MnCl$_2$ and ZnCl$_2$ exhibited unexpected behavior, where both the paramagnetic and diamagnetic ions migrated toward regions of higher magnetic field gradients. This result contrasts with the behavior of single ions and suggests that ionic interactions between the two metal species lead to the formation of mixed, field-induced clusters. The stronger magnetic response of MnCl$_2$ appeared to dominate the overall migration dynamics, driving clusters that contains both MnCl$_2$ and ZnCl$_2$ toward the magnet surface. Additionally, the concentration ratio between the two metal ions played a critical role. Increasing the initial concentration of diamagnetic ZnCl$_2$ in the mixture, reduced the extent of MnCl$_2$ enrichment, highlighting competitive effects likely driven by complex ion-ion interactions and opposing magnetic forces.\par 

In Part II of this study, we will conduct a more rigorous analysis of these experimental findings by developing a comprehensive multi-physics theoretical framework. This framework will incorporate detailed numerical simulations to quantitatively interpret the observed magnetophoresis phenomena and provide deeper insights into the governing mechanisms.\par 

\section{Acknowledgments}
A portion of this work was performed at the National High Magnetic Field Laboratory, which is supported by the National Science Foundation Cooperative Agreement No. DMR-1644779 and the state of Florida. This work was supported by the Center for Rare Earths, Critical Minerals, and Industrial Byproducts, through funding provided by the State of Florida. HM acknowledges the support from National Science Foundation through award CET 2343151.
\bibliography{myrefs}

\end{document}